\documentclass[amsmath,amssymb,reqno,tbtags,psamsfonts,10pt,a4paper,twocolumn,showpacs, prl]{revtex4}

\usepackage[english]{babel}
\usepackage{stmaryrd}

\usepackage{color,graphicx}
\usepackage{upgreek}
\usepackage[normalem]{ulem} % enables the crossing out of sentences with the command \sout{}
\usepackage{epstopdf} % allows you to use eps figures and still compile with pdflatex

\usepackage{mathptmx} % changes font to Times, including math
\newcommand{\Fig}[1]{Fig.~\ref{#1}}

\begin{document}

\title[]{Localization dynamics of fluids in random confinement}
\bigskip
\author{Thomas O. E. Skinner$^{1}$, Simon K. Schnyder$^{2}$, Dirk G. A. L. Aarts$^1$, 
J{\"u}rgen Horbach$^{2}\footnote{e-mail: horbach@thphy.uni-duesseldorf.de}$,
and Roel P. A. Dullens$^{1}\footnote{e-mail: roel.dullens@chem.ox.ac.uk}$}
\affiliation{$^1$Department of Chemistry, Physical and Theoretical Chemistry Laboratory, 
                 University of Oxford, South Parks Road, Oxford OX1 3QZ, United Kingdom,\\
            $^2$Institut f\"{u}r Theoretische Physik II, Heinrich-Heine-Universit\"{a}t 
                 D\"{u}sseldorf, Universit\"{a}tsstra{\ss}e 1, 40225 D\"{u}sseldorf, Germany}

\pacs{82.70.Dd, 66.30.H-, 64.60.Ht, 61.43.-j}

\begin{abstract}
The dynamics of two-dimensional fluids confined within a random matrix of
obstacles is investigated using both colloidal model experiments and molecular
dynamics simulations.  By varying fluid and matrix area fractions in the
experiment, we find delocalized tracer particle dynamics at small matrix area 
fractions and localized motion of the tracers at high matrix area fractions. 
In the delocalized region, the dynamics is subdiffusive at intermediate times, and diffusive at long times, while in the localized regime, trapping in finite pockets of the matrix is observed.
These observations are found to agree
with the simulation of an ideal gas confined in a weakly correlated matrix.
Our results show that Lorentz gas systems with soft interactions are exhibiting a smoothening of the critical dynamics and consequently a rounded delocalization-to-localization transition. 
\end{abstract}

\maketitle

{\it Introduction.---}Understanding the dynamics in
disordered heterogeneous media is of great interest for fields like materials science, geophysics or biology
\cite{sahimi1993,dingwell96,hofling2013}. The mass transport in
such media is associated with a strong separation of time scales,
i.e.~mobile particles that move through a
``matrix of immobile'' particles.  Experimental realizations of
such systems are binary mixtures of colloids with disparate sizes
\cite{Imhof1995,Imhof1995a}, ion-conducting glasses
\cite{jonscher77,bunde98,horbach02,Voigtmann2006a,dyre09} or biological
systems like cells \cite{hofling2013}.

To model such heterogeneous systems one often treats the matrix particles
as fixed. One of the simplest models is
the disordered Lorentz gas \cite{Lorentz1905,Beijeren1982} where a tracer particle moves in a matrix of fixed, randomly distributed, overlapping hard spheres. This model exhibits a
delocalization-to-localization transition associated with the
critical percolation point \cite{Stauffer1991} of the void space. At low matrix densities
the void space is a percolating network leading to diffusive tracer motion, but it becomes disconnected above the critical percolation density and the tracer is trapped in
finite pockets of void space. At the critical point, the long-time diffusion is anomalous
\cite{Ben-Avraham2000,Hofling2006,Bauer2010}, characterized by a sublinear
dependence of the mean squared displacement (\textsc{msd}) on time , $\delta r^2(t) \sim t^x$
with $x\approx 2/3.036$ in two dimensions (2D) \cite{Bauer2010} and 
$x\approx 0.32$ in 3D \cite{Hofling2006}.

It is an open question as to whether the Lorentz gas scenario is also relevant in complex media
sucha as ion conductors or porous materials. Simulations have reported anomalous 
diffusion and localization dynamics in various more realistic heterogeneous
systems such as binary mixtures with a frozen-in component
\cite{Kurzidim2009,Kurzidim2010,Kurzidim2011,Kim2009,Kim2010,Kim2011}
or a disparate size ratio \cite{Voigtmann2009} and
realistic models of ion-conducting alkali silicate glasses
\cite{horbach02}.  These studies are corroborated by
mode coupling theory calculations \cite{goetze_book}, which provide a qualitative
description of the Lorentz gas \cite{Spanner2013} and predict localization transitions for disparate-sized
binary mixtures \cite{Voigtmann2011} and quenched-annealed systems
\cite{Krakoviack2005,Krakoviack2007,Krakoviack2009,Krakoviack2011}.

While models like the Lorentz gas are insightful, experimental
realizations of systems with fixed, randomly placed matrix
particles are very rare. Most experimental studies of diffusion in disordered media have been reported for porous glasses
\cite{Bellissent1993} where systematically varying the
pore size distribution is very difficult.
In this Letter, we present a 2D colloidal model experiment 
consisting of small tracer particles in a random confining matrix of fixed large
particles. The experimental setup enables \emph{in situ} control of the area
fractions of the fluid and matrix particles \emph{without} changing the matrix
configuration, which is crucial for studying the localization dynamics in random media. The experiment exhibits delocalized tracer dynamics at low matrix area fractions and highly localized tracer motion at high matrix area fractions. Our experimental
findings are supported by molecular dynamics (\textsc{md})
simulations of a 2D soft sphere binary mixture 
with the large species being fixed and the small species forming an ideal fluid 
of noninteracting particles. The simulations indicate that the experimental localization
transition is rounded due to the soft interactions, which lead to a distribution of energies of the tracers and finite barriers in the matrix. This is qualitatively different from the sharp
localization transition in the Lorentz gas.

\begin{figure}[htbp]
\includegraphics[width=\columnwidth]{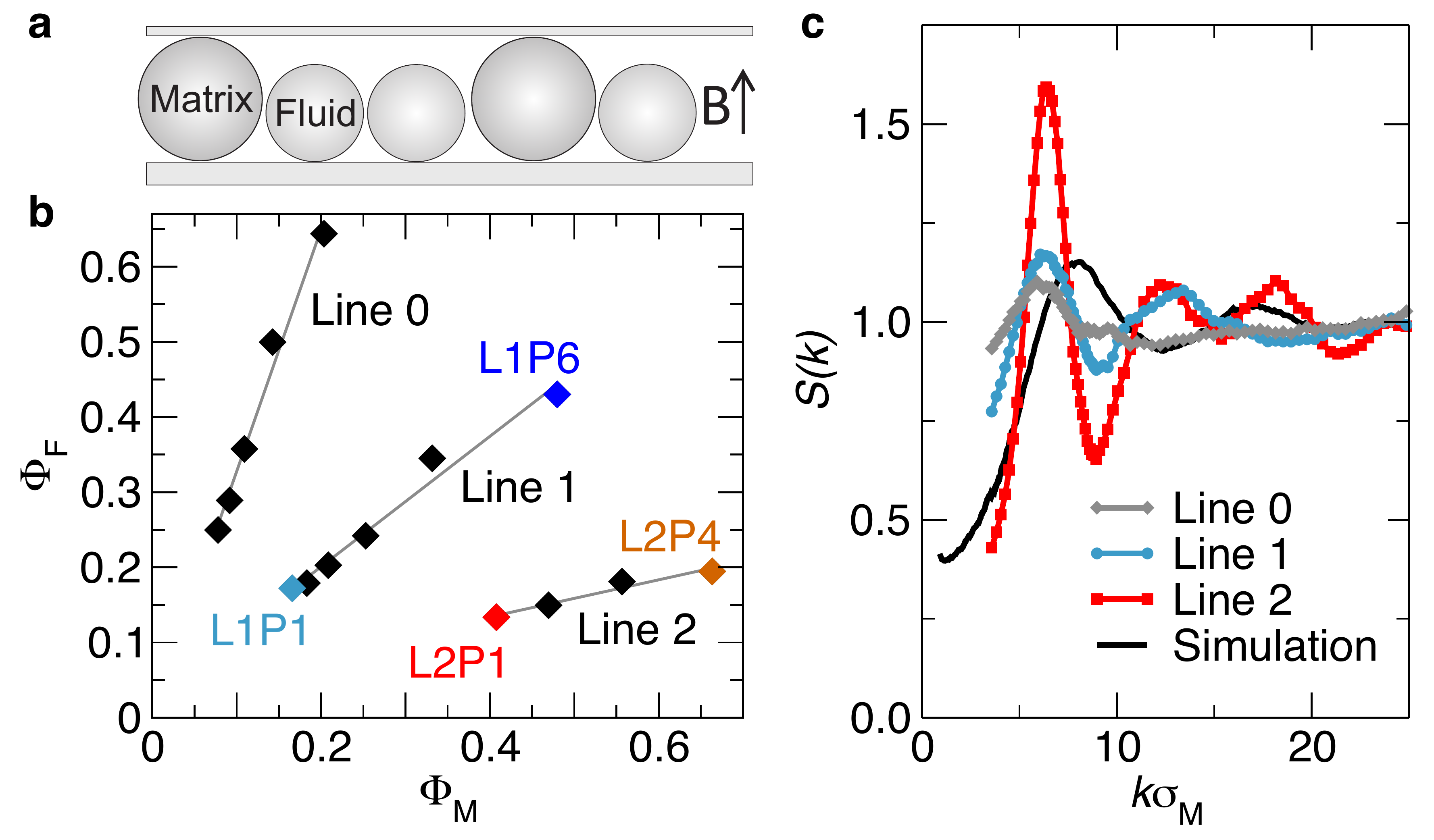}
\caption{(color online). a) The sample cell with the matrix
particles acting as spacers. b) State diagram for the effective area fractions
of the fluid ($\Phi_{\mathrm F}$) versus the matrix particles ($\Phi_{\mathrm
M}$). c) Structure factors of the matrix configurations for lines 0, 1 and 2 and the simulation.} \label{state_d}
\end{figure}

{\it Experiment.---}A mixture of 3.9\,$\upmu$m and
4.95\,$\upmu$m diameter superparamagnetic colloidal polystyrene spheres
(Microparticles GmbH) in water is confined between two glass
slides to make a 2D sample cell (\Fig{state_d}a). The large particles act as spacers and hence form the fixed matrix in which the small particles -- the fluid -- are free to move \cite{CruzdeLeon1998}. Video microscopy is 
used to image an area of 430\,$\upmu$m$~\times$ 340\,$\upmu$m for up to 2\,h. Standard tracking routines are used to find the matrix
and fluid particle coordinates \cite{Crocker1996} as a function of time
from which we computed the static and dynamics quantities of interest.

In the experiments, we prepared three different samples which
are represented by the three lines denoted line 0, 1 and 2 in the state diagram in \Fig{state_d}b. Each line, i.e. sample, corresponds to a different matrix configuration and the lowest state point along each line is characterized by the hard sphere area fractions of the matrix ($\Phi_{\mathrm
M}^0$) and the fluid ($\Phi_{\mathrm F}^0$) particles and is thus achieved
without the presence of a magnetic field. The lowest state point of each line is labeled as `L1P1' for
line 1 for example. To achieve higher \emph{effective} packing fractions
$\Phi_{\mathrm{F,M}}$, we apply a perpendicular external magnetic field $B$
which leads to a repulsive pair potential given by
%${U_{\mathrm{F,M}}(r)=\frac{\mu_{0} \chi_{\mathrm{F,M}}^2 B^2}{4\pi r^3}}$.
 ${U_{\mathrm{F,M}}(r)=\mu_{0} \chi_{\mathrm {F,M}}^2 B^2/ \left ( 4\pi r^3 \right )}$.  
Here, $r$ is the distance between two particles, $\mu_0$ the 
permeability of free space and $\chi_{\mathrm{F,M}}$ the magnetic susceptibility
of the fluid or matrix particles. The state points with the magnetic field present are e.g. `L1P2, ..., L1P6' for line 1. To
represent these state points in the $(\Phi_{\mathrm{M}},\Phi_{\mathrm{F}})$
state diagram, we calculate the effective hard sphere diameters
$\sigma_{\mathrm{F,M}}$ using a Barker-Henderson approach:
${\sigma_{\mathrm {F,M}}=\sigma^0_{\mathrm {F,M}}
+ \int_{\sigma^0_{\mathrm {F,M}}}^{\infty} \left ( 1 - \mathrm{e}^{- \beta U_\mathrm{F,M}(r)} \right ) \mathrm{d}r}$, where $\sigma^{0}_{\mathrm{F,M}}$ are the hard sphere diameters and $\beta =
1/k_{B}T$ \cite{Barker1967,Hansen2006}. If $B=0$, $\sigma_{\mathrm{F,M}}$ reduces to $\sigma_{\mathrm{F,M}}^0$, which corresponds
to the lowest state point along each line. Note that the effective
matrix to fluid particle size ratio remains fairly constant at 0.787 as is
evident from the linear paths in the $(\Phi_{\mathrm{F}},\Phi_{\mathrm{M}})$
state diagram (\Fig{state_d}b).

The key point of our approach is that we \emph{in situ} control 
the effective area fractions \emph{without} changing the matrix configuration, so that we can very efficiently probe the tracer dynamics at different effective matrix and fluid area fractions. Importantly, we retain the random character of the matrix even at very high
effective area fractions, which is crucial for studying the localization
dynamics in random media and comparing it to the Lorentz gas. In our analysis of the tracer dynamics, we will focus on the state points along lines 1 and 2. The experimental data are averaged over up to five independent matrix configurations by imaging different parts of each sample.

First, we characterize the structural correlations of the matrix by
computing its static structure factor ${S(k)=\frac{1}{N_M} \left
\langle \sum_{i=1}^{N_{M}} \sum_{j=1}^{N_{M}} \mathrm{e}^{-i \vec{k} \cdot
(\vec{r}_i - \vec{r}_j)} \right \rangle}$. Here, $N_{\rm M}$ is the number of
matrix particles, the angled brackets represent an ensemble average and
$\vec{r}_i$ and $\vec{r}_j$ are the positions of matrix particles $i$ and
$j$, respectively. The matrix structure factors corresponding
to lines 0, 1 and 2 only show weak fluidlike structural correlation
(\Fig{state_d}c), similar to the uncorrelated Lorentz model. Note that
the structure factors do \emph{not} change along each
line, which is a direct consequence of increasing the
effective area fractions by increasing $B$. Importantly, it also directly confirms that the matrix particles are fixed.

\begin{figure*}[tb]
\includegraphics[width=\textwidth]{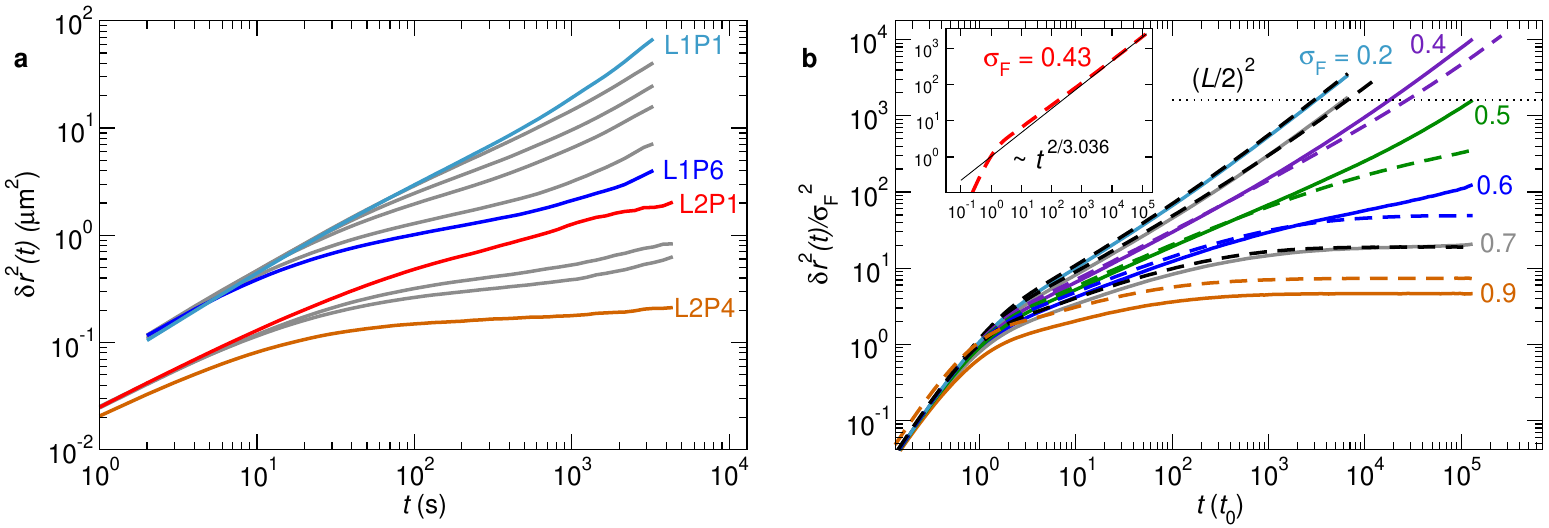}
\caption{(color online). a) \textsc{msd} for the fluid particles in the
experiments along lines 1 and 2.  b) \textsc{msd} for
the fluid particles in the simulation for $\sigma_\mathrm{F} = 0.2$, 0.3, 0.4, 0.5, 0.6, 0.7, and 0.9 for
mono-energetic tracers (dashed lines) and tracers with a distribution of energies (solid lines). The
dotted line roughly marks the limit above which finite size effects
become important (with simulation box size $L$). 
Inset: Comparison of the \textsc{msd} of the mono-energetic system at
$\sigma_\mathrm{F}=0.43$ to the expected critical asymptote of the
two-dimensional Lorentz model.}
\label{msd}
\end{figure*}

We analyze the fluid particle dynamics by computing the mean squared displacement ${\delta r^2 (t)
= \frac{1}{N_{F}} \left< \sum_{i=1}^{N_{ F}} [\vec{r}_i(t) - \vec{r}_i(0)]^2
\right>}$, with $N_{\rm F}$ the number of fluid particles and $\vec{r}_i$
the position of fluid particle $i$. The \textsc{msd}s along lines 1
and 2 are shown in \Fig{msd}a. Note that the difference in the short-time diffusion is due to the different number densities
\cite{qiu1990, peng2008} and that diffusion is well defined in 2D systems with fixed obstacles \cite{Bauer2010}. At L1P1, characterized by low matrix and
low fluid area fractions, the tracer dynamics is diffusive at long
times. Upon increasing the effective area fractions $\Phi_{\rm F}$ and $\Phi_{\rm M}$
along line 1, the dynamics shows a noticeable slowing down and strongly
subdiffusive behavior at intermediate times. However, at long times the
\textsc{msd}s of all the state points along line 1 exhibit diffusive
behavior, corresponding to delocalized motion. Consistently, delocalized motion is observed at all state 
points along line 0, i.e., at lower matrix area fractions (data not shown). In contrast, the fluid particle
dynamics exhibit completely different behavior along line 2, i.e., at much higher
matrix area fractions. At L2P1, the \textsc{msd} plateaus at long
times, which upon increasing $\Phi_{\rm{F}}$ and $\Phi_{\rm{M}}$
along line 2, decreases to smaller values. This behavior clearly indicates the
localization of the tracers in this region of the state diagram and
shows that the length scale associated with the localization becomes smaller as
the matrix area fraction is increased. 

{\it Simulation.---}To gain more insight into the mechanism of the experimentally 
observed localized tracer dynamics, we perform detailed \textsc{md}
simulations of a quenched-annealed system: a binary system of purely repulsive
soft spheres with the Weeks-Chandler-Andersen interaction
potential \cite{weeks1971}, where the matrix particles are fixed. In particular,
we simulate the tracer dynamics in the limit of $\Phi_{\rm F} \rightarrow 0$,
while we systematically increase the matrix area fraction, which enables
us to efficiently sample the regions corresponding to both delocalized and 
localized tracer dynamics. 
Importantly, we do not aim to achieve quantitative agreement between the experiments and the simulations. This is prohibitively difficult as -- among other things -- one would have to include hydrodynamic interactions and account for the change of the softness of the potential with the magnetic field. Instead, we feel that it is far more instructive to perform simulations with a different soft interaction potential and hence reveal generic features of the localization dynamics in Lorentz gas systems with soft interactions.

The parameters of the system and the \textsc{md} simulation are set as in
\cite{Voigtmann2009}, except that the tracer interaction is turned off to avoid
the effect of fluid-fluid interaction on the dynamics. The matrix particle
diameters are sampled equidistantly $\sigma_{\mathrm M}\in[0.85,1.15]$ to
avoid crystallization. For  each realization of the system, between 1000 and
4000 matrix particles are equilibrated at number density
$\rho=0.278\sigma_\mathrm{M}^{-2}$ at the temperature $k_\mathrm{B}T=1$ by
randomly selecting their velocities from the Maxwell distribution every 100
steps, for $10^5$ time steps.  Then, the matrix particles are fixed and their
positions are uniformly rescaled to $\rho=0.625\sigma_\mathrm{M}^{-2}$. The
computed quantities are averaged over 100 independent matrix configurations. To
simulate the tracer dynamics we insert up to 1000 noninteracting fluid
particles into each matrix configuration. We vary the tracer diameter
$\sigma_\mathrm{F}$, which is equivalent to varying the matrix area fraction
\emph{without} changing the structure of the matrix, analogous to the
experiments. Varying the system size $L$ allows us to control finite size
effects, which only start to play a role after the \textsc{msd} exceeds
$\approx (L/2)^2$ as indicated in \Fig{msd}b.

\begin{figure*}[tbh]
\centering
\includegraphics[width=1.01\columnwidth]{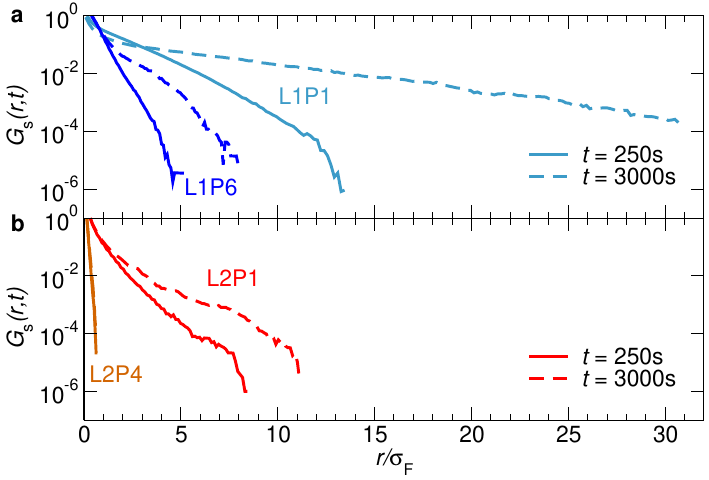}
\hfill
\includegraphics[width=1.01\columnwidth]{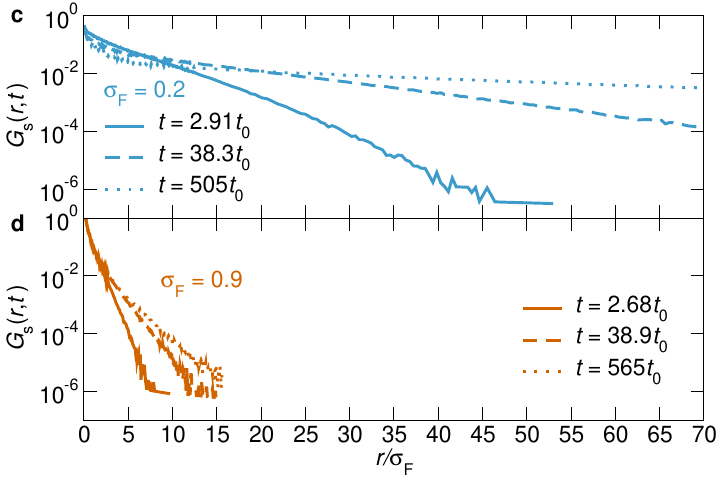}
\caption{(color online). a) $G_{\mathrm s}(r,t)$ for the fluid
particles in the experiment along line 1 at state points 1 and 2 after $250$s 
and $3000$s.
b) As a), but for line 2.  c) $G_{\mathrm s}(r,t)$ for the fluid particles in the simulation for a delocalized state ($\sigma_{\mathrm{F}}=0.2$), d) As  c), but for a localized state ($\sigma_{\mathrm{F}}=0.9$).}
\label{van_Hove}
\end{figure*}

{\it Comparison between simulation and experiment.---}As in the experiments, the structure factor of the simulated matrix only shows weak fluidlike structural correlations (\Fig{state_d}c). 
In the simulations, we consider two cases: the tracers have either the \emph{same} or a \emph{distribution} of energies. 
In the first case, the tracers are given the same energy by first determining the average energy during equilibration and then re-inserting them such that the energy of each tracer equals the average energy. The simulations then reproduce the Lorentz model, as is evident from the \textsc{msd} of the tracers for different values of $\sigma_{\rm F}$ (dashed lines in  \Fig{msd}b). In the long-time limit, the \textsc{msd} changes from diffusive
at small $\sigma_{\rm F}$ to localized at large $\sigma_{\rm F}$ with an
extended subdiffusive regime at intermediate times. The
localization-delocalization transition occurs approximately at $\sigma_{\rm F}^*
= 0.43$ (see the inset), where the \textsc{msd} asymptotically approaches a power law,
$\delta r^2 \sim t^x$ with $x\approx 2/3.036$, in agreement with the Lorentz
model \cite{Bauer2010}. 

The experimental system, however, is \emph{not} an ensemble of tracers with
the same energy but exhibits a broad distribution of tracer energies. Moreover, the barriers between the pores in the matrix
are of finite height due to the soft interactions, which allows tracers
with high energy to overcome barriers that cannot be surpassed by low-energy tracers. 
As a consequence, tracers with different energies have different critical points, which necessarily leads
to an averaging of the dynamics: the localization transition is expected to be
rounded and hence the Lorentz model exponent cannot be measured in our experiments. To demonstrate this effect, we consider the second case and  simulate a system where the tracers form an ideal gas, which is the
simplest system of tracers with a distribution of energies. For comparison to the single-energy case, the average energy per tracer
particle is kept the same. The ideal gas \textsc{msd}s (solid lines in \Fig{msd}b) clearly differ from the single-energy case (dashed lines) at long times. The difference is most striking at $\sigma_{\rm F}=0.6$,
where the single-energy system is clearly localized, while the ideal gas system
still shows diffusion at long times, as a substantial subset of the
particles has high enough energies to be delocalized.

Our simulations thus show that Lorentz gas systems with soft interactions -- 
as our experimental system -- exhibit a rounded
delocalization-to-localization transition. This can also be inferred from the
experimental data in Fig.~\ref{msd}a: the tracers at state point
L1P6 are still diffusive at long times, while at state point L2P1, which has a
comparable $\Phi_{\rm M}$ but a far lower $\Phi_{\rm F}$, the tracers are
localized. This is a direct consequence of the potential at high
magnetic fields (L1P6) being much softer, which leads to a rounding of the
localization transition, whereas L2P1 is characterized by a far harder potential
and less softening of the localization transition. 

To further corroborate our findings, we also compute the self-part of
the van Hove correlation function $G_{\mathrm s}(r,t)$ as it gives the full spatial
information for the tracer particle dynamics at a given time. This
function gives the distribution of displacements $\Delta r_i = \left|
\vec{r}_i(t) - \vec{r}_i(0)\right|$ of a tagged particle $i$ at time $t$:
$G_{\mathrm{s}}(r,t)= \frac{1}{N_{F}} \left<
\sum_{i=1}^{N_{F}} \delta \left(r - \Delta r_i\right) \right>$ \cite{glassbook}.  In
\Fig{van_Hove}a, we show the experimental $G_{\mathrm s}(r,t)$ for L1P1 and L1P6 at times $t=250$\,s and $t=3000$\,s. 
The lower time (250\,s, solid lines) roughly corresponds to the time where
subdiffusion is strongest, while the larger time (3000\,s, dashed
lines) marks the end of the experimental runs. The delocalized nature of the 
tracers at low area fractions is reflected by the broad distribution at
L1P1 and its time dependence: the width increases
by a factor of 3 from $t=250$\,s to $t=3000$\,s. At point L1P6
the shape of $G_{\mathrm{s}}(r,t)$ is very similar compared to L1P1
but the distributions are narrower at both times reflecting the much higher 
effective area fractions. The $G_{\mathrm s}(r,t)$'s for L2P1 and L2P4
are shown in \Fig{van_Hove}b. Although the fluid particles at L2P1 are localized, which can only be inferred
here from the time dependence i.e., from the \textsc{msd}, the
$G_{\mathrm{s}}(r,t)$ is fairly similar to that of L1P6 -- despite the fact
that L2P1 is characterized by a far lower fluid area fraction. 
The very narrow $G_{\mathrm{s}}(r,t)$ and the absence of any shift when
changing from $t=250$\,s to $t=3000$\,s indicates a strong localization at L2P4.

In Figs.~\ref{van_Hove}c and \ref{van_Hove}d we show $G_{\mathrm s}(r,t)$ for the simulated confined
ideal gas at three times, where the values of the \textsc{msd}
roughly agree with those of the experiment: $t\approx 3 t_0$,
$t\approx39 t_0$ and an additional time that is order of magnitude larger, 
$t\approx 550 t_0$.  The shape of $G_\mathrm{s}(r,t)$
of the simulation at $\sigma_{\mathrm{F}}=0.2$ (\Fig{van_Hove}c),
matches the $G_\mathrm{s}(r,t)$ of L1P1 (\Fig{van_Hove}a) well. As
$\sigma_\mathrm{F}$ increases, $G_\mathrm{s}(r,t)$ in the simulation
undergoes the same qualitative development as in the experiment.
On the localized side ($\sigma_{\mathrm{F}}=0.9$, \Fig{van_Hove}d), a very 
narrow distribution is observed that shows virtually no change with time, except from a small broadening which
qualitatively matches the experimental data at L2P1 and L2P4. The general agreement between experiment and simulation also shows that
fluid-fluid interactions are not important in the experiment. This is additionally indicated by the fact that there is no sign of particle hopping in the experimental $G_\mathrm{s}(r,t)$ which would  in peaks or
shoulders at distances comparable to the nearest-neighbor distance between two tracer particles (note that the quantity $2 \pi r G_\mathrm{s}(r,t)$ also does not show
such a feature) \cite{horbach02,Kurzidim2010}.

{\it Conclusion.---}We have studied the localization dynamics of two-dimensional 
fluids confined in a random matrix using colloidal experiments and molecular dynamics simulations. In the experiments, mean squared
displacements and van Hove correlation functions show signatures of delocalized 
tracer dynamics at low matrix area fractions and localized motion at high matrix
area fractions. In particular, we observe long-time diffusion at
small matrix area fractions, while trapping in finite pockets of the matrix is 
present at high matrix area fractions. 
The molecular dynamics simulations show that the
soft interactions in our colloidal model system, which give rise to an energy distribution
for the tracer particles, smoothen the critical dynamics and enhance the
diffusivity, leading to a rounded localization transition. Our
results show that the smoothening of the critical dynamics does not depend on
the details of the interaction potential, which suggests that the rounding of
the localization transition is a generic feature of realistic systems.

\begin{acknowledgments}
We thank Pinaki Chaudhuri for useful discussions.
S.K.S. and J.H. acknowledge the \textsc{dfg} research unit FOR-1394 ``Nonlinear 
response to probe vitrification'' (HO 2231/7-1), T.O.E.S., 
D.G.A.L.A. and R.P.A.D. the EPSRC; J.H. and R.P.A.D. the Royal Society for financial support.
\end{acknowledgments}

% \bibliographystyle{apsrev-mod}
% \bibliography{rand_conf_letter}

\begin{thebibliography}{39}
\expandafter\ifx\csname natexlab\endcsname\relax\def\natexlab#1{#1}\fi
\expandafter\ifx\csname bibnamefont\endcsname\relax
  \def\bibnamefont#1{#1}\fi
\expandafter\ifx\csname bibfnamefont\endcsname\relax
  \def\bibfnamefont#1{#1}\fi
\expandafter\ifx\csname citenamefont\endcsname\relax
  \def\citenamefont#1{#1}\fi
\expandafter\ifx\csname url\endcsname\relax
  \def\url#1{\texttt{#1}}\fi
\expandafter\ifx\csname urlprefix\endcsname\relax\def\urlprefix{URL }\fi
\providecommand{\bibinfo}[2]{#2}
\providecommand{\eprint}[2][]{\url{#2}}

\bibitem[{\citenamefont{Sahimi}(1993)}]{sahimi1993}
\bibinfo{author}{\bibfnamefont{M.}~\bibnamefont{Sahimi}},
  \bibinfo{journal}{Rev. Mod. Phys.} \textbf{\bibinfo{volume}{65}},
  \bibinfo{pages}{1393} (\bibinfo{year}{1993}).

\bibitem[{\citenamefont{Dingwell}(1996)}]{dingwell96}
\bibinfo{author}{\bibfnamefont{D.~B.} \bibnamefont{Dingwell}},
  \bibinfo{journal}{Science} \textbf{\bibinfo{volume}{273}},
  \bibinfo{pages}{1054} (\bibinfo{year}{1996}).

\bibitem[{\citenamefont{H\"{o}fling and Franosch}(2013)}]{hofling2013}
\bibinfo{author}{\bibfnamefont{F.}~\bibnamefont{H\"{o}fling}}
\bibnamefont{and}
  \bibinfo{author}{\bibfnamefont{T.}~\bibnamefont{Franosch}},
  \bibinfo{journal}{Rep. Prog. Phys.} \textbf{\bibinfo{volume}{76}} \bibinfo{pages}{046602}  (\bibinfo{year}{2013}).

\bibitem[{\citenamefont{Imhof and Dhont}(1995{\natexlab{a}})}]{Imhof1995}
\bibinfo{author}{\bibfnamefont{A.}~\bibnamefont{Imhof}} \bibnamefont{and}
  \bibinfo{author}{\bibfnamefont{J.~K.~G.} \bibnamefont{Dhont}},
  \bibinfo{journal}{Phys. Rev. E} \textbf{\bibinfo{volume}{52}},
  \bibinfo{pages}{6344} (\bibinfo{year}{1995}{\natexlab{a}}).

\bibitem[{\citenamefont{Imhof and Dhont}(1995{\natexlab{b}})}]{Imhof1995a}
\bibinfo{author}{\bibfnamefont{A.}~\bibnamefont{Imhof}} \bibnamefont{and}
  \bibinfo{author}{\bibfnamefont{J.~K.~G.} \bibnamefont{Dhont}},
  \bibinfo{journal}{Phys. Rev. Lett.} \textbf{\bibinfo{volume}{75}},
  \bibinfo{pages}{1662} (\bibinfo{year}{1995}{\natexlab{b}}).

\bibitem[{\citenamefont{Jonscher}(1977)}]{jonscher77}
\bibinfo{author}{\bibfnamefont{A.~K.} \bibnamefont{Jonscher}},
  \bibinfo{journal}{Nature} \textbf{\bibinfo{volume}{267}},
  \bibinfo{pages}{673} (\bibinfo{year}{1977}).

\bibitem[{\citenamefont{Bunde et~al.}(1998)\citenamefont{Bunde, Funke, and
  Ingram}}]{bunde98}
\bibinfo{author}{\bibfnamefont{A.}~\bibnamefont{Bunde}},
  \bibinfo{author}{\bibfnamefont{K.}~\bibnamefont{Funke}}, \bibnamefont{and}
  \bibinfo{author}{\bibfnamefont{M.~D.} \bibnamefont{Ingram}},
  \bibinfo{journal}{Solid State Ionics} \textbf{\bibinfo{volume}{105}},
  \bibinfo{pages}{1} (\bibinfo{year}{1998}).

\bibitem[{\citenamefont{Horbach et~al.}(2002)\citenamefont{Horbach, Kob, and
  Binder}}]{horbach02}
\bibinfo{author}{\bibfnamefont{J.}~\bibnamefont{Horbach}},
  \bibinfo{author}{\bibfnamefont{W.}~\bibnamefont{Kob}}, \bibnamefont{and}
  \bibinfo{author}{\bibfnamefont{K.}~\bibnamefont{Binder}},
  \bibinfo{journal}{Phys. Rev. Lett.} \textbf{\bibinfo{volume}{88}},
  \bibinfo{pages}{125502} (\bibinfo{year}{2002}).

\bibitem[{\citenamefont{Voigtmann and Horbach}(2006)}]{Voigtmann2006a}
\bibinfo{author}{\bibfnamefont{T.}~\bibnamefont{Voigtmann}} \bibnamefont{and}
  \bibinfo{author}{\bibfnamefont{J.}~\bibnamefont{Horbach}},
  \bibinfo{journal}{EPL} \textbf{\bibinfo{volume}{74}}, \bibinfo{pages}{459}
  (\bibinfo{year}{2006}).

\bibitem[{\citenamefont{Dyre et~al.}(2009)\citenamefont{Dyre, Maass, Roling,
  and Sidebottom}}]{dyre09}
\bibinfo{author}{\bibfnamefont{J.~C.} \bibnamefont{Dyre}},
  \bibinfo{author}{\bibfnamefont{P.}~\bibnamefont{Maass}},
  \bibinfo{author}{\bibfnamefont{B.}~\bibnamefont{Roling}}, \bibnamefont{and}
  \bibinfo{author}{\bibfnamefont{D.~L.} \bibnamefont{Sidebottom}},
  \bibinfo{journal}{Rep. Prog. Phys.} \textbf{\bibinfo{volume}{72}},
  \bibinfo{pages}{046501} (\bibinfo{year}{2009}).

\bibitem[{\citenamefont{Lorentz}(1905)}]{Lorentz1905}
\bibinfo{author}{\bibfnamefont{H.~A.} \bibnamefont{Lorentz}},
  \bibinfo{journal}{Proc. R. Acad. Sci. Amsterdam} \textbf{\bibinfo{volume}{7}},
  \bibinfo{pages}{438} (\bibinfo{year}{1905}).

\bibitem[{\citenamefont{van Beijeren}(1982)}]{Beijeren1982}
\bibinfo{author}{\bibfnamefont{H.}~\bibnamefont{van Beijeren}},
  \bibinfo{journal}{Rev. Mod. Phys.} \textbf{\bibinfo{volume}{54}},
  \bibinfo{pages}{195} (\bibinfo{year}{1982}).

\bibitem[{\citenamefont{Stauffer and Aharony}(1991)}]{Stauffer1991}
\bibinfo{author}{\bibfnamefont{D.}~\bibnamefont{Stauffer}} \bibnamefont{and}
  \bibinfo{author}{\bibfnamefont{A.}~\bibnamefont{Aharony}},
  \emph{\bibinfo{title}{{Introduction to Percolation Theory}}}
  (\bibinfo{publisher}{Taylor and Francis}, \bibinfo{address}{London},
  \bibinfo{year}{1991}).

\bibitem[{\citenamefont{Ben-Avraham and Havlin}(2000)}]{Ben-Avraham2000}
\bibinfo{author}{\bibfnamefont{D.}~\bibnamefont{Ben-Avraham}}
\bibnamefont{and}
  \bibinfo{author}{\bibfnamefont{S.}~\bibnamefont{Havlin}},
  \emph{\bibinfo{title}{{Diffusion and Reactions in Fractals and Disordered
  Systems}}} (\bibinfo{publisher}{Cambridge University Press},
  \bibinfo{address}{Cambridge, England}, \bibinfo{year}{2000}).

\bibitem[{\citenamefont{H\"{o}fling et~al.}(2006)\citenamefont{H\"{o}fling,
  Franosch, and Frey}}]{Hofling2006}
\bibinfo{author}{\bibfnamefont{F.}~\bibnamefont{H\"{o}fling}},
  \bibinfo{author}{\bibfnamefont{T.}~\bibnamefont{Franosch}},
\bibnamefont{and}
  \bibinfo{author}{\bibfnamefont{E.}~\bibnamefont{Frey}},
  \bibinfo{journal}{Phys. Rev. Lett.} \textbf{\bibinfo{volume}{96}},
  \bibinfo{pages}{165901} (\bibinfo{year}{2006}).

\bibitem[{\citenamefont{Bauer et~al.}(2010)\citenamefont{Bauer, H\"{o}fling,
  Munk, Frey, and Franosch}}]{Bauer2010}
\bibinfo{author}{\bibfnamefont{T.}~\bibnamefont{Bauer}},
  \bibinfo{author}{\bibfnamefont{F.}~\bibnamefont{H\"{o}fling}},
  \bibinfo{author}{\bibfnamefont{T.}~\bibnamefont{Munk}},
  \bibinfo{author}{\bibfnamefont{E.}~\bibnamefont{Frey}}, \bibnamefont{and}
  \bibinfo{author}{\bibfnamefont{T.}~\bibnamefont{Franosch}},
  \bibinfo{journal}{Eur. Phys. J. Special Topics} \textbf{\bibinfo{volume}{189}},
  \bibinfo{pages}{103} (\bibinfo{year}{2010}).

\bibitem[{\citenamefont{Kurzidim et~al.}(2009)\citenamefont{Kurzidim,
  Coslovich, and Kahl}}]{Kurzidim2009}
\bibinfo{author}{\bibfnamefont{J.}~\bibnamefont{Kurzidim}},
  \bibinfo{author}{\bibfnamefont{D.}~\bibnamefont{Coslovich}},
  \bibnamefont{and} \bibinfo{author}{\bibfnamefont{G.}~\bibnamefont{Kahl}},
  \bibinfo{journal}{Phys. Rev. Lett.} \textbf{\bibinfo{volume}{103}},
  \bibinfo{pages}{138303} (\bibinfo{year}{2009}).

\bibitem[{\citenamefont{Kurzidim et~al.}(2010)\citenamefont{Kurzidim,
  Coslovich, and Kahl}}]{Kurzidim2010}
\bibinfo{author}{\bibfnamefont{J.}~\bibnamefont{Kurzidim}},
  \bibinfo{author}{\bibfnamefont{D.}~\bibnamefont{Coslovich}},
  \bibnamefont{and} \bibinfo{author}{\bibfnamefont{G.}~\bibnamefont{Kahl}},
  \bibinfo{journal}{Phys. Rev. E} \textbf{\bibinfo{volume}{82}},
  \bibinfo{pages}{041505} (\bibinfo{year}{2010}).

\bibitem[{\citenamefont{Kurzidim et~al.}(2011)\citenamefont{Kurzidim,
  Coslovich, and Kahl}}]{Kurzidim2011}
\bibinfo{author}{\bibfnamefont{J.}~\bibnamefont{Kurzidim}},
  \bibinfo{author}{\bibfnamefont{D.}~\bibnamefont{Coslovich}},
  \bibnamefont{and} \bibinfo{author}{\bibfnamefont{G.}~\bibnamefont{Kahl}},
  \bibinfo{journal}{J. Phys. Condens. Matter} \textbf{\bibinfo{volume}{23}},
  \bibinfo{pages}{234122} (\bibinfo{year}{2011}).

\bibitem[{\citenamefont{Kim et~al.}(2009)\citenamefont{Kim, Miyazaki, and
  Saito}}]{Kim2009}
\bibinfo{author}{\bibfnamefont{K.}~\bibnamefont{Kim}},
  \bibinfo{author}{\bibfnamefont{K.}~\bibnamefont{Miyazaki}},
\bibnamefont{and}
  \bibinfo{author}{\bibfnamefont{S.}~\bibnamefont{Saito}},
  \bibinfo{journal}{Europhys. Lett.} \textbf{\bibinfo{volume}{88}}, \bibinfo{pages}{36002}
  (\bibinfo{year}{2009}).

\bibitem[{\citenamefont{Kim et~al.}(2010)\citenamefont{Kim, Miyazaki, and
  Saito}}]{Kim2010}
\bibinfo{author}{\bibfnamefont{K.}~\bibnamefont{Kim}},
  \bibinfo{author}{\bibfnamefont{K.}~\bibnamefont{Miyazaki}},
\bibnamefont{and}
  \bibinfo{author}{\bibfnamefont{S.}~\bibnamefont{Saito}},
  \bibinfo{journal}{Eur. Phys. J. Special Topics} \textbf{\bibinfo{volume}{189}},
  \bibinfo{pages}{135} (\bibinfo{year}{2010}).

\bibitem[{\citenamefont{Kim et~al.}(2011)\citenamefont{Kim, Miyazaki, and
  Saito}}]{Kim2011}
\bibinfo{author}{\bibfnamefont{K.}~\bibnamefont{Kim}},
  \bibinfo{author}{\bibfnamefont{K.}~\bibnamefont{Miyazaki}},
\bibnamefont{and}
  \bibinfo{author}{\bibfnamefont{S.}~\bibnamefont{Saito}},
\bibinfo{journal}{J.
  Phys. Condens. Matter} \textbf{\bibinfo{volume}{23}},
  \bibinfo{pages}{234123} (\bibinfo{year}{2011}).

\bibitem[{\citenamefont{Voigtmann and Horbach}(2009)}]{Voigtmann2009}
\bibinfo{author}{\bibfnamefont{T.}~\bibnamefont{Voigtmann}} \bibnamefont{and}
  \bibinfo{author}{\bibfnamefont{J.}~\bibnamefont{Horbach}},
  \bibinfo{journal}{Phys. Rev. Lett.} \textbf{\bibinfo{volume}{103}},
  \bibinfo{pages}{205901} (\bibinfo{year}{2009}).

\bibitem[{\citenamefont{G\"otze}(2009)}]{goetze_book}
\bibinfo{author}{\bibfnamefont{W.}~\bibnamefont{G\"otze}},
  \emph{\bibinfo{title}{{Complex Dynamics in Glass-Forming Liquids}}}
  (\bibinfo{publisher}{Oxford University Press}, \bibinfo{address}{Oxford, England},
  \bibinfo{year}{2009}).

\bibitem[{\citenamefont{Spanner et~al.}(2013)\citenamefont{Spanner, Schnyder,
  H\"{o}fling, Voigtmann, and Franosch}}]{Spanner2013}
\bibinfo{author}{\bibfnamefont{M.}~\bibnamefont{Spanner}},
  \bibinfo{author}{\bibfnamefont{S.~K.} \bibnamefont{Schnyder}},
  \bibinfo{author}{\bibfnamefont{F.}~\bibnamefont{H\"{o}fling}},
  \bibinfo{author}{\bibfnamefont{T.}~\bibnamefont{Voigtmann}},
  \bibnamefont{and}
\bibinfo{author}{\bibfnamefont{T.}~\bibnamefont{Franosch}},
  \bibinfo{journal}{Soft Matter} \textbf{\bibinfo{volume}{9}},
  \bibinfo{pages}{1604} (\bibinfo{year}{2013}).

\bibitem[{\citenamefont{Voigtmann}(2011)}]{Voigtmann2011}
\bibinfo{author}{\bibfnamefont{T.}~\bibnamefont{Voigtmann}},
  \bibinfo{journal}{Europhys. Lett.} \textbf{\bibinfo{volume}{96}}, \bibinfo{pages}{36006}
  (\bibinfo{year}{2011}).

\bibitem[{\citenamefont{Krakoviack}(2005)}]{Krakoviack2005}
\bibinfo{author}{\bibfnamefont{V.}~\bibnamefont{Krakoviack}},
  \bibinfo{journal}{Phys. Rev. Lett.} \textbf{\bibinfo{volume}{94}},
  \bibinfo{pages}{065703} (\bibinfo{year}{2005}).

\bibitem[{\citenamefont{Krakoviack}(2007)}]{Krakoviack2007}
\bibinfo{author}{\bibfnamefont{V.}~\bibnamefont{Krakoviack}},
  \bibinfo{journal}{Phys. Rev. E} \textbf{\bibinfo{volume}{75}},
  \bibinfo{pages}{031503} (\bibinfo{year}{2007}).

\bibitem[{\citenamefont{Krakoviack}(2009)}]{Krakoviack2009}
\bibinfo{author}{\bibfnamefont{V.}~\bibnamefont{Krakoviack}},
  \bibinfo{journal}{Phys. Rev. E} \textbf{\bibinfo{volume}{79}},
  \bibinfo{pages}{061501} (\bibinfo{year}{2009}).

\bibitem[{\citenamefont{Krakoviack}(2011)}]{Krakoviack2011}
\bibinfo{author}{\bibfnamefont{V.}~\bibnamefont{Krakoviack}},
  \bibinfo{journal}{Phys. Rev. E} \textbf{\bibinfo{volume}{84}},
  \bibinfo{pages}{050501} (\bibinfo{year}{2011}).

\bibitem[{\citenamefont{Bellissent-funel
  et~al.}(1993)\citenamefont{Bellissent-funel, Lal, and
  Bosio}}]{Bellissent1993}
\bibinfo{author}{\bibfnamefont{M.~C.} \bibnamefont{Bellissent-Funel}},
  \bibinfo{author}{\bibfnamefont{J.}~\bibnamefont{Lal}}, \bibnamefont{and}
  \bibinfo{author}{\bibfnamefont{L.}~\bibnamefont{Bosio}},
\bibinfo{journal}{J.
  Chem. Phys.} \textbf{\bibinfo{volume}{98}}, \bibinfo{pages}{4246}
  (\bibinfo{year}{1993}).

\bibitem[{\citenamefont{{Cruz de Le\'{o}n} et~al.}(1998)\citenamefont{{Cruz de
  Le\'{o}n}, Saucedo-Solorio, and Arauz-Lara}}]{CruzdeLeon1998}
\bibinfo{author}{\bibfnamefont{G.}~\bibnamefont{{Cruz de Le\'{o}n}}},
  \bibinfo{author}{\bibfnamefont{J.~M.}~\bibnamefont{Saucedo-Solorio}},
  \bibnamefont{and}
  \bibinfo{author}{\bibfnamefont{J.~L.}~\bibnamefont{Arauz-Lara}},
  \bibinfo{journal}{Phys. Rev. Lett.} \textbf{\bibinfo{volume}{81}},
  \bibinfo{pages}{1122} (\bibinfo{year}{1998}).

\bibitem[{\citenamefont{Crocker and Grier}(1996)}]{Crocker1996}
\bibinfo{author}{\bibfnamefont{J.~C.} \bibnamefont{Crocker}} \bibnamefont{and}
  \bibinfo{author}{\bibfnamefont{D.~G.} \bibnamefont{Grier}},
  \bibinfo{journal}{J. Colloid Interface Sci.} \textbf{\bibinfo{volume}{179}},
  \bibinfo{pages}{298} (\bibinfo{year}{1996}).

\bibitem[{\citenamefont{Barker and Henderson}(1967)}]{Barker1967}
\bibinfo{author}{\bibfnamefont{J.~A.} \bibnamefont{Barker}} \bibnamefont{and}
  \bibinfo{author}{\bibfnamefont{D.}~\bibnamefont{Henderson}},
  \bibinfo{journal}{J. Chem. Phys.} \textbf{\bibinfo{volume}{47}},
  \bibinfo{pages}{4714} (\bibinfo{year}{1967}).

\bibitem[{\citenamefont{Hansen and McDonald}(2006)}]{Hansen2006}
\bibinfo{author}{\bibfnamefont{J.~P.} \bibnamefont{Hansen}} \bibnamefont{and}
  \bibinfo{author}{\bibfnamefont{I.~R.} \bibnamefont{McDonald}},
  \emph{\bibinfo{title}{{Theory of simple liquids}}}
  (\bibinfo{publisher}{Academic Press}, \bibinfo{address}{London},
  \bibinfo{year}{2006}).

\bibitem[{\citenamefont{Qiu et~al.}(1990)\citenamefont{Qiu, Wu, Xue, Pine,
  Weitz, and Chaikin}}]{qiu1990}
\bibinfo{author}{\bibfnamefont{X.}~\bibnamefont{Qiu}},
  \bibinfo{author}{\bibfnamefont{X.~L.} \bibnamefont{Wu}},
  \bibinfo{author}{\bibfnamefont{J.~Z.} \bibnamefont{Xue}},
  \bibinfo{author}{\bibfnamefont{D.~J.} \bibnamefont{Pine}},
  \bibinfo{author}{\bibfnamefont{D.~A.} \bibnamefont{Weitz}},
\bibnamefont{and}
  \bibinfo{author}{\bibfnamefont{P.~M.} \bibnamefont{Chaikin}},
  \bibinfo{journal}{Phys. Rev. Lett.} \textbf{\bibinfo{volume}{65}},
  \bibinfo{pages}{516} (\bibinfo{year}{1990}).

\bibitem[{\citenamefont{Peng et~al.}(2008)\citenamefont{Peng, Chen, Fischer,
  Weitz, and Tong}}]{peng2008}
\bibinfo{author}{\bibfnamefont{Y.}~\bibnamefont{Peng}},
  \bibinfo{author}{\bibfnamefont{W.}~\bibnamefont{Chen}},
  \bibinfo{author}{\bibfnamefont{T.~M.} \bibnamefont{Fischer}},
  \bibinfo{author}{\bibfnamefont{D.~A.} \bibnamefont{Weitz}},
\bibnamefont{and}
  \bibinfo{author}{\bibfnamefont{P.}~\bibnamefont{Tong}}, \bibinfo{journal}{J.
  Fluid Mech.} \textbf{\bibinfo{volume}{618}}, \bibinfo{pages}{243}
  (\bibinfo{year}{2009}).

\bibitem[{\citenamefont{Weeks et~al.}(1971)\citenamefont{Weeks, Chandler, and
  Andersen}}]{weeks1971}
\bibinfo{author}{\bibfnamefont{J.~D.} \bibnamefont{Weeks}},
  \bibinfo{author}{\bibfnamefont{D.}~\bibnamefont{Chandler}}, \bibnamefont{and}
  \bibinfo{author}{\bibfnamefont{H.~C.} \bibnamefont{Andersen}},
  \bibinfo{journal}{J. Chem. Phys.} \textbf{\bibinfo{volume}{54}},
  \bibinfo{pages}{5237} (\bibinfo{year}{1971}).

\bibitem[{\citenamefont{Binder and Kob}(2011)}]{glassbook}
\bibinfo{author}{\bibfnamefont{K.}~\bibnamefont{Binder}} \bibnamefont{and}
  \bibinfo{author}{\bibfnamefont{W.}~\bibnamefont{Kob}},
  \emph{\bibinfo{title}{{Glassy Materials and Disordered Solids}}}
  (\bibinfo{publisher}{World Scientific}, \bibinfo{address}{Singapore},
  \bibinfo{year}{2011}).

\end{thebibliography}
\end{document}